\documentclass[12pt]{article}

\usepackage{times,epsfig,graphics,latexsym,amssymb,latexsym,
amsmath,setspace,appendix,epstopdf}
\usepackage{array,threeparttable,hyperref,url}
\usepackage[usenames]{color}
\usepackage[authoryear,sort]{natbib}
\usepackage{booktabs}
\usepackage{dcolumn}
\newcolumntype{d}{D{.}{.}{-1}}
\allowdisplaybreaks

\newtheorem{thm}{Theorem}
\newtheorem{lem}{Lemma}

\newtheorem{rmk}{Remark}

\def\log{\hbox{log}}

\def\boxit#1{\vbox{\hrule\hbox{\vrule\kern6pt
          \vbox{\kern6pt#1\kern6pt}\kern6pt\vrule}\hrule}}

\def\bse{\begin{eqnarray*}}
\def\ese{\end{eqnarray*}}
\def\be{\begin{eqnarray}}
\def\ee{\end{eqnarray}}
\def\bq{\begin{equation}}
\def\eq{\end{equation}}

\newtheorem{proposition}{Proposition}

\newcommand{\blem}{\begin{lemma}}
\newcommand{\elem}{\end{lemma}}
\newcommand{\bthe}{\begin{theorem}}
\newcommand{\ethe}{\end{theorem}}

\newtheorem{definition}{Definition}[section]
\newtheorem{lemma}[definition]{Lemma}
\newtheorem{theorem}[definition]{Theorem}

\def\delete#1{\iffalse #1 \fi}

\def\bse{\begin{eqnarray*}}
\def\ese{\end{eqnarray*}}
\def\bee{\begin{enumerate}}
\def\eee{\end{enumerate}}
\def\bqe{\begin{eqnarray}}
\def\eqe{\end{eqnarray}}
\def\bed{\begin{description}}
\def\eed{\end{description}}
\def\bei{\begin{itemize}}
\def\eei{\end{itemize}}

\def\pmb#1{\setbox0=\hbox{#1}%
    \kern-.025em\copy0\kern-\wd0
    \kern.05em\copy0\kern-\wd0
    \kern-.025em\raise.0433em\box0 }
\def\pmbh#1#2{\setbox0=\hbox{#1}%
    \setbox1=\hbox{#2}%
    \kern-.025em\copy0\kern-\wd0
    \kern.05em\copy1\kern-\wd0
    \kern-.025em\raise.0433em\box0 }

\def\frac#1#2{{#1\over#2}}

\def\boxit#1{\vbox{\hrule\hbox{\vrule\kern6pt
   \vbox{\kern6pt#1\kern6pt}\kern6pt\vrule}\hrule}}

\def\listing#1{\vskip 4mm\begin{verbatim}\input#1 \vskip 4mm}
\def\thick#1{\hbox{\rlap{$#1$}\kern0.25pt\rlap{$#1$}\kern0.25pt$#1$}}

\def\pmbh{{\pmb h}}

\def\calF{{\cal F}}

\def\calJ{{\cal J}}

\def\calM{{\cal M}}
\def\calN{{\cal N}}

\def\calT{{\cal T}}

\renewcommand\today{\ifcase\month\or
   Jan\or Feb\or Mar\or Apr\or May\or
   Jun\or Jul\or Aug\or Sep\or Oct\or Nov\or
   Dec\fi
   \space\number\day, \number\year}

\newcommand{\vu}{{\bf u}}

\newcommand{\vx}{{\bf x}}
\newcommand{\vy}{{\bf y}}

\newcommand{\vI}{{\bf I}}

\newcommand{\vP}{{\bf P}}
\newcommand{\vQ}{{\bf Q}}

\newcommand{\vX}{{\bf X}}

\newcommand{\vnull}{{\bf 0}}

\newcommand{\vbeta}{\mbox{\boldmath $\beta$}}

\newcommand{\bay}{\begin{array}}
\newcommand{\eay}{\end{array}}

\newcommand{\bqa}{\begin{eqnarray*}}
\newcommand{\eqa}{\end{eqnarray*}}
\newcommand{\bqan}{\begin{eqnarray}}
\newcommand{\eqan}{\end{eqnarray}}
\newcommand{\bqt}{\begin{quote}}
\newcommand{\eqt}{\end{quote}}
\newcommand{\bt}{\begin{tabbing}}
\newcommand{\et}{\end{tabbing}}
\newcommand{\bit}{\begin{itemize}}
\newcommand{\eit}{\end{itemize}}
\newcommand{\ben}{\begin{enumerate}}
\newcommand{\een}{\end{enumerate}}
\newcommand{\beq}{\begin{equation}}
\newcommand{\eeq}{\end{equation}}
\newcommand{\bdefi}{\begin{definition}}
\newcommand{\edefi}{\end{definition}}
\newcommand{\bpro}{\begin{proposition}}
\newcommand{\epro}{\end{proposition}}
\newcommand{\bco}{\begin{corollary}}
\newcommand{\eco}{\end{corollary}}
\newcommand{\bdes}{\begin{description}}
\newcommand{\edes}{\end{description}}

\newcommand{\vepsilon}{\mbox{\boldmath $\epsilon$}}

\DeclareMathOperator*{\argmin}{arg\,min}

\def\boxit#1{\vbox{\hrule\hbox{\vrule\kern6pt\vbox{\kern6pt#1\kern6pt}\kern6pt\vrule}\hrule}}

\catcode`@=11
\def\@evenhead{\vbox{\hbox to\textwidth{\tiny \hfill \hfill \today } }}
\def\@oddhead{\vbox{\hbox to \textwidth{\tiny \hfill \hfill \today } }}

\def\author.arg{

Ming-Yen Cheng, Sanying Feng, Gaorong Li, and Heng Lian

}

\def\tit.arg{Greedy Forward Regression for Variable Screening}

\begin{document}
\pagenumbering{arabic}
\setcounter{page}{1}

\baselineskip=18pt

\begin{center}
{\Large \tit.arg}
\footnote{
Ming-Yen Cheng is Professor, Department of Mathematics, National Taiwan University, Taipei 106, Taiwan (email: cheng@math.ntu.edu.tw).  
Sanying Feng is Lecturer, School of Mathematics and Statistics, Zhengzhou University, Zhengzhou 450001, China (email: fsy5801@emails.bjut.edu.cn). 
Gaorong Li is Associate Professor, Beijing Institute for Scientific and Engineering Computing, Beijing University of Technology, Beijing  100124,  China (email: ligaorong@bjut.edu.cn).
Heng Lian is Senior Lecturer, School of Mathematics and Statistics, University of New South Wales, Sydney, NSW 2052, Australia (email: heng.lian@unsw.edu.au). 
Cheng's research was supported by the Ministry of Science and Technology grant 
104-2118-M-002-005-MY3. Feng's research was supported by the National Natural Science Foundation of China (11501522).
Li's research was supported by the National Natural Science Foundation of China (11471029),
the Beijing Natural Science Foundation (1142002), the Science and Technology Project of
 Beijing Municipal Education Commission (KM201410005010) and Program for JingHua Talents in
  Beijing University of Technology (2013-JH-L07).}
\vskip 3mm

\author.arg
\end{center}

\vskip 3mm

\date{}

\begin{abstract}
Two popular variable screening methods under the ultra-high dimensional setting with the desirable sure screening property are the sure independence screening (SIS) and the forward regression (FR). Both are classical variable screening methods and recently have attracted greater attention under the new light of high-dimensional data analysis. We consider a new and simple screening method that incorporates multiple predictors in each step of forward regression, with decision on which variables to incorporate based on the same criterion. If only one step is carried out, it actually reduces to the SIS. Thus it can be regarded as a generalization and unification of the FR and the SIS. More importantly, it preserves the sure screening property and has similar computational complexity as FR in each step, yet it can discover the relevant covariates in fewer steps. Thus, it reduces the computational burden of FR drastically while retaining advantages of the latter over SIS. Furthermore, we show that it can find all the true variables if the number of steps taken is the same as the correct model size, even when using the original FR. An extensive simulation study and application to two real data examples demonstrate excellent performance of the proposed method.
 \end{abstract}

\vskip .1 in
\noindent {{\bf Key words}: \it Bayesian information criterion; Independence Screening; Model selection / Variable selection; Ultra-high dimensionality.
}

\newpage

\section{Introduction}
With rapid advances of modern technology, datasets involving a huge number of variables arise frequently from biological, business, financial, genetics, social studies, etc. Mathematically speaking, we often need to deal with ultra-high dimensional statistical problems by which we  mean that $\log (p)$ can be as large as $n^a$ for some constant $a\in (0,1)$, where $p$ and $n$ denote the dimension and sample size, respectively. For example, in the two genetics examples analyzed in Section \ref{sec:app}, $p$ is in tens of thousands whereas $n$ is only around one hundred. Here we focus on multiple linear regression. Extensions to other parametric or semiparametric models are possible but would require additional notation and technical treatments which would obscure the ideas. 

In dealing with high-dimensional problems, sparsity is a typical assumption in order to reduce the effective number of parameters and to make estimation feasible.
Various penalized regression methods have been proposed for simultaneous selection and estimation under sparsity assumptions. \cite{tibshirani96} proposed lasso, whose theoretical properties are investigated by numerous works including \cite{knightfu00,zhaoyu06,zhanghuang08,bickel09}. Following its success, many different penalty functions have been proposed to deal with the known issues of lasso, and these methods have been extended to more general regression models. \cite{fan01} and \cite{zou06} proposed SCAD penalty and adaptive lasso penalty, respectively, which are consistent in variable selection. \cite{zouhastie05} suggested elastic net by combining lasso and ridge penalties, which can better deal with collinearity in covariates.  \cite{fan04}, \cite{kim2008smoothly}, \cite{huangma08ss}, \cite{zouzhang09}, and \cite{fan2011non} investigated theoretical properties of these penalties when the dimension diverges or grows faster than the sample size. \cite{fanlv2013} characterized the asymptotic equivalence of different regularization methods. \cite{fanli2012} studied regularized estimation in linear mixed effects model. In semiparametric additive or varying coefficient models, penalized estimation has been considered by \cite{xuequzhou10,huang10,wei11,liansinica12,wangxuequliang14}, among many others. Such methods have become a standard approach to high-dimensional or big data analysis conducted in a diverse spectrum of research fields.

Despite the popularity and success of these penalized methods, they may not perform well due to the ``challenges of computational expediency, statistical accuracy and algorithmic stability'' \citep{fansamworthwu09,fansong10}. In particular, computational efficiency is a major concern when the dimension is more than, say, a thousand or even more, because complex optimization algorithms are often used in these methods.
To cope with these problems, \cite{fan2008sure} proposed a sure independence screening (SIS) method to screen out unimportant variables and reduce the dimensionality to a manageable order. Specifically, SIS achieves independence screening by ranking marginal correlations between individual covariates and the response variable. The surprising theoretical result is that this simple procedure possesses the sure screening property, that is, all the predictors in the true model with nonzero coefficients will be included in the estimated model, under mild assumptions. Motivated by its favorable performance and its ease of use in practice, many have followed the lead of \cite{fan2008sure} and proposed various ways to improve its performance and to extend it to different models. \cite{fansong10} generalized SIS to generalized linear models.
\cite{zhulilizhu11} and \cite{lizhongzhu12} proposed model-free screening without parametric assumption based on sufficient dimension reduction and distance correlation, respectively. \cite{lipengzhangzhu12} proposed a robust screening procedure based on Kendall $\tau$'s rank correlation. \cite{fanfengsong11,fanmadai14,liuliwu14, CHLP2014,SYZ2014} considered independence screening for semiparametric additive and varying-coefficient models. \cite{hewanghong13} studied variable screening in quantile regression for both parametric and semiparametric models. \cite{FKLZ2015} introduced interaction screening for nonlinear classification. 

Soon after the proposal of SIS, \cite{wang09fr} showed that another popular and classical variable selection method, namely the forward regression (FR), also possesses the sure screening property in sparse ultra-high dimensional linear models. Although he did not claim in the work that FR is the only good method for variable screening, the numerical simulations demonstrated the superior performance of FR. In particular, while FR and SIS have similar coverage probabilities the former has a much lower false discovery rate than the latter. This may be due to the fact that FR at least partially takes into account the correlations among covariates by performing multiple linear regression using all the currently incorporated variables, while SIS ignores the effects of all the other covariates when computing the marginal correlation. This also shows up in the technical assumptions required in demonstrating their sure screening property. Specifically, FR only requires the coefficients in the true linear model are sufficiently large (see our definition of $\beta_{\min}$ given in Theorem \ref{thm:1}) i.e. the sparsity assumption. By comparison, SIS typically requires the marginal correlations of the relevant covariates with the response are sufficiently high, which is in general not true even if the coefficients
in the true linear model are large. On the other hand, due to the necessity to perform multiple linear regression, FR is certainly slower to compute than SIS. Thus, it would be helpful to reduce the number of steps in FR while keeping its superior properties at the same time.

Motivated by the above mentioned observations, we propose an extension of FR, called greedy forward regression (GFR). It incorporates multiple covariates, say $J$ of them, into the estimated model in each step of the sequential selection. Note that doing this directly by choosing additional $J$ variables that reduce the sum of squares of residuals (SSR) the most would cause extra computational burden, defeating the computational expediency of using variable screening methods. The reason is this approach would require performing and comparing ${p\choose J}$ regression models in each step. Instead, the key idea of our proposal is that we still compute the reduction in SSR when adding only one variable each time and only in the last step we will incorporate multiple variables. 

In our theoretical study we show the sure screening property of the proposed GFR method. We also show another stronger theoretical property which is new even when $J=1$ i.e. when the original FR is used. Specifically, we study the number of steps required to find all the true important variables, and we show that all of them will be identified if we fix the number of steps the same as the true model size. 
Compared to the standard forward regression, the theoretical challenge here is further caused by the fact that in GFR we compute the reduction in SSR by adding one variable while we include multiple variables in each step. Therefore, our theoretical results are non-trivial. Selection of the tuning parameter $J$ is a minor issue and we provide some general suggestions. Our numerical studies demonstrate that the proposed method retains the advantages of FR over SIS (and ISIS, an iterative variant of SIS) while improving on FR in terms of computational speed.

The details of our algorithm are contained in Section \ref{sec:gfr}, and the theoretical results are given in Section \ref{sec:theory} . In Section \ref{sec:sim} we examine the finite sample performance and compare with the traditional FR, SIS and ISIS via an extensive simulation study. Section \ref{sec:app} presents and discusses application of our method to two genetics datasets. Given in Section \ref{sec:con} are conclusions and future studies. Proofs of the theoretical results are deferred to Section \ref{sec:prof}.

\section{Greedy Forward Regression}\label{sec:gfr}
We consider standard linear regression models.
Let $(\vX_1,y_1),\ldots,(\vX_n,y_n)$ be a sample of independent observations obeying the following form:
$$y_i=\mu+\sum_{j=1}^pX_{ij}\beta_j+\epsilon_i,$$
where $\vX_i=(X_{i1},\ldots,X_{ip})^T$ is the $p$-dimensional covariate vector of the $i$th observation, $\mu$ is the intercept, $\vbeta=(\beta_1,\ldots,\beta_p)^T$ are the unknown coefficients and $\epsilon_i$ is the mean zero error contained in the $i$th observation. In the rest of the paper we assume $\mu=0$ for simplicity of notation. In matrix notation, we write
$$\vy=\vX\vbeta+\vepsilon,$$
with $\vy=(y_1,\ldots,y_n)^T$, $\vX=(\vX_1,\ldots,\vX_p)_{n\times p}$ and $\vepsilon=(\epsilon_1,\ldots,\epsilon_n)^T$. We focus on the ultra-high dimensional regime where $p>>n$ and assume a sparse true model in which $p_0:=|\{j:\beta_j\neq 0\}|$ is smaller than $n$. The true model is denoted by $\calT=\{j:\beta_j\neq 0\}$ while the full model is written as $\calF=\{1,\ldots,p\}$. For any submodel $\calM\subseteq\calF$, let $\vX_\calM$ be the submatrix of $\vX$ associated with $\calM$; it has $|\calM|$ columns. Similarly $\vbeta_\calM$ denotes the subvector of $\vbeta$ containing only components in $\calM$.

In SIS, we rank the importance of variables by $|\vX_j^T\vy|$, $j\in \calF$. It only requires going through each of the $p$ predictors once; therefore SIS is computationally expeditious. However, in general $\beta_j\neq 0$ in the true model does not imply $|\vX_j^T\vy|\neq 0$ in the marginal model. Thus SIS directly assumes $|\vX_j^T\vy|\neq 0$ when $\beta_j\neq 0$ in order to guarantee its sure screening property. 
In the FR algorithm, starting with the null model, we incorporate variables into the model one at a time. At each step, every variable that is not already in the current model is tested and the one which  reduces the sum of squares of residuals (SSR) the most is added to the model. Since fitting the submodel in each step is necessary, FR is computationally slower than SIS although it has better control on the false discovery rate. One obvious modification of FR is to consider the best set of $J$ variables that reduce SSR the most if added to the current model together. However, this would increase considerably the computational burden in each step  because there are ${p-|\calM| \choose J}$ possible ways to pick up $J$ variables out of the $p-|\calM|$ candidates, if $\calM$ is the current model.
To avoid this computational problem, in each step of our greedy FR algorithm, we still compute the SSR for each variable outside the current model, and pick the $J$ variables that reduce SSR the most marginally. The algorithm is more formally presented below. In the following we use $\vP_{\calM}=\vX_{\calM}(\vX_{\calM}^T\vX_{\calM})^{+}\vX_{\calM}$ for the projection matrix associated with $span\{\vX_{\calM}\}$, the column span of $\vX_{\calM}$, where $()^+$ denotes the Moore-Penrose pseudo-inverse. Let $\vQ_{\calM}=\vI-\vP_{\calM}$ be the projection to the subspace orthogonal to $span\{\vX_{\calM}\}$.

\noindent {\bf Greedy forward regression algorithm:} 
\begin{itemize}
\item[(i)] Choose the tuning parameter $J\ge 1$. Initially we start with the null model $\calM^{(0)}=\emptyset$, and set the step number $k$ as $k=1$.
\item[(ii)] In step $k$, let $\calN^{(k)}=\{j_1,\ldots,j_J\}$ be the index set of the predictors such that the values of $\|\vP_{\calM^{(k-1)}\cup\{j\}}\vy\|, j\in\calN^{(k)}$, are  the $J$ largest among all those $j\in \calF\backslash\calM^{(k-1)}$. Set $\calM^{(k)}=\calM^{(k-1)}\cup\calN^{(k)}$.
\item[(iii)] Repeat (ii) until at least $n-J+1$ covariates are incorporated (with more than $n$ covariates the least squares problem becomes unidentified).
\end{itemize}
Obviously $\|\vP_{\calM^{(k-1)}\cup\{j\}}\vy\|^2=\|\vy\|^2-\|\vQ_{\calM^{(k-1)}\cup\{j\}}\vy\|^2$ and the procedure in (ii) is the same as choosing the predictors associated with the $J$ smallest values of $\|\vQ_{\calM^{(k-1)}\cup\{j\}}\vy\|^2$, i.e. the SSR for model $\calM^{(k-1)}\cup\{j\}$, among all $j\in \calF\backslash\calM^{(k-1)}$.

Similar to forward regression, even with $J>1$, in each step we perform at most $p$ projections $\vP_{\calM^{(k-1)}\cup\{j\}}\vy$ (or $\vQ_{\calM^{(k-1)}\cup\{j\}}\vy$), $j\in \calF\backslash\calM^{(k-1)}$. However, we need to keep track of the $J$ largest values along the way, which would incur some extra computational burden. Empirically, we find that such additional book-keeping only adds a small amount of computational time to the algorithm. Detailed computational time comparisons are made in our simulation studies.

Finally, we note that if we set $J$ to be large, say $J=n$ or $n/\log n$ and perform the procedure in (ii) only once, then our algorithm reduces to the marginal independence screening of \cite{fan2008sure}. Thus GFR can be regarded as an extension of both FR and independence screening, as taking $J=1$ it corresponds to FR and choosing $J$ close to $n$ it reduces to SIS. On the other hand, empirically we suggest to choose relatively small value of $J$ such as 2 or 4. Thus GFR builds a bridge between FR and SIS. In the numerical studies we examine its finite sample performance and find that in general it is superior to both FR and SIS.

\section{Theoretical Properties}\label{sec:theory}
Since more than one predictor is added to the model in each step of the greedy FR algorithm, certainly it will take fewer steps to reach a model with a target model size. For example, as in (iii) of the algorithm we stop as soon as at least $n-J+1$ predictors are incorporated. However, we are more interested in how this approach affects the consistency of the screening algorithm. Suppose the algorithm builds a finite sequence of models $\calM^{(1)},\ldots,\calM^{(K)}$, usually referred to as the solution path. We say the algorithm produces a consistent solution path in variable screening if
$$P(\calT\subseteq\calM^{(k)} \mbox{ for some }   k )\rightarrow 1.$$
This definition was used by \cite{wang09fr} for the original forward regression.

For greedy FR with $J>1$, the first question is of course whether it still has the desirable screening consistency property. A more refined question is regarding the smallest value of $k$ such that $\calT\subseteq\calM^{(k)}$. That is, how many iterations are needed before all the relevant predictors are included in the model? Comparing greedy FR with the original FR, intuitively, the worst case that can happen is the additional $J-1$ covariates selected in each step are not relevant at all and the number of iterations required is the same as that required by FR. The best case, on the other hand, is that all the additional $J-1$ covariates included in each step are ``as relevant as" the top one and the number of iterations is thus reduced by a factor of $J$. Our first result, given in Theorem \ref{thm:1}, shows that the best case happens at least in the upper bound we obtain for the number of iterations required by the greedy FR. Thus, it has the potential to incorporate all the relevant predictors in fewer steps than FR does.
Our second theoretical result given in Theorem \ref{thm:2} tries to answer the same question, but from a slightly different perspective. We consider the following question: under what conditions will the greedy FR with $J\geq 1$ incorporate at least one relevant predictor in each step? When this happens, it will incorporate all relevant predictors after at most $p_0$ steps. It turns out this happens under reasonable assumptions. This result appears to be new even for the case of $J=1$, i.e. the FR, to our knowledge. 

We first define restricted eigenvalues and restricted correlations, which have been used for example in \cite{bickel09}.
For an integer $s$, the restricted eigenvalues are defined as
$$\phi(s)=\min_{\|\vx\|_0\le s}\frac{\vx^T\vX^T\vX\vx}{n\|\vx\|^2} \qquad \mbox{and} \qquad
\Phi(s)=\max_{\|\vx\|_0\le s}\frac{\vx^T\vX^T\vX\vx}{n\|\vx\|^2},$$
and the restricted correlations are
$$\theta_{s_1,s_2}=\max\left\{\frac{\vx_1^T\vX_{\calM_1}^T\vX_{\calM_2}\vx_2}{n\|\vx_1\|\|\vx_2\|}: \calM_1\cap\calM_2=\emptyset, |\calM_1|\le s_1,|\calM_2|\le s_2\right\}.$$
In particular, by definition, we have $\|\vX_j\|^2\le n\Phi(1)$, where $\vX_j$ is the $j$-th column of $\vX$.
In some literature, it is assumed that $\phi(s)$ and $\Phi(s)$ are bounded and bounded away from zero for $s=O(n^{\alpha})$ with some value $\alpha<1$, which will simplify the bounds below. We choose to explicitly track  these quantities for the sake of generality, and only require they are nonzero.

\begin{thm}\label{thm:1} Assume $\epsilon_i$ has a subgaussian distribution. That is, there exists a constant $c>0$ such that $E[\exp\{t\epsilon_i\}]\le \exp\{ct^2\}$. Let $\beta_{\min}=\min_{j\in \calT} |\beta_j|$.
Suppose $K_0$ is an integer satisfying
\begin{equation}\label{eqn:cond1}
K_0> \frac{2\|\vy\|^2\Phi(J)\Phi(1)}{n\phi^3(p_0K_0J)J\beta_{\min}^2}
\end{equation}
and
\begin{equation*}
p_0K_0J \log(p)=o_p(\frac{n\phi^2(p_0K_0J)\beta_{\min}^2}{\Phi(1)}),
\end{equation*}
then $$P(\calT\subseteq\calM^{(p_0K_0)})\rightarrow 1.$$
That is, all relevant variables are incorporated after $p_0K_0$ steps.
\end{thm}
\begin{rmk}
Suppose that for a constant $C$ sufficiently large, $\tau_1<\phi(s)\le\Phi(s)\le\tau_2$ for two positive constants when $s\le Cp_0/\beta_{\min}^2$. If we further assume reasonably that $\|\vy\|^2=O_p(n)$, then $K_0$ can be chosen to be $K_0=O_p(1/(J\beta_{\min}^2))$ and at most $O_p(p_0/(J\beta_{\min}^2))$ steps are required. When $J=1$, this gives a $O_p(p_0/\beta_{\min}^2)$ bound on the number steps required, which is better than the $O_p(p_0^2/\beta_{\min}^4)$ bound stated in Theorem 1 of \cite{wang09fr}. The reason for the improvement here is that we use a slightly tighter lower bound in (\ref{eqn:qmq2}) of the proof, compared to their equation (B.6).
\end{rmk}
\begin{rmk} Some assumptions are implicit in the statement of the theorem above. These include $\phi(p_0K_0J)\neq 0$ and $\beta_{\min}\neq 0$. Also implicitly assumed is $p_0K_0\le [n/J]$, where $[n/J]$ denotes the integer part of $n/J$, since we will terminate the algorithm after $[n/J]$ iterations.
\end{rmk}
\begin{rmk} From the proof it can be seen that the constant 2 in (\ref{eqn:cond1}) can be replaced be any fixed constant larger than 1.
\end{rmk}
From Theorem \ref{thm:1}, it is seen that using $J>1$ the greedy FR algorithm will discover all the relevant predictors in fewer steps. However, the trade-off is that each step of greedy FR incorporates $J$ covariates which makes the computation slower when comparing the $k$-th step of greedy FR with that of the original FR. Although the theorem seems to suggest that a larger value of $J$ is better, we note that it merely provides an upper bound on the number of iterations required. Another hidden condition is that, since $p_0K_0J\approx n$ and $K_0\ge 1$, we need $J\le n/p_0$. Empirically, we find a relatively small value of $J$, say $J=2$ or 4, works better.

\begin{thm}\label{thm:2}
Assume the noises are subgaussian. Suppose for some $\eta>0$,
\begin{equation}\label{eqn:thm2cond10}
\frac{\phi^3(p_0J)J}{\Phi(1)p_0}\ge (1+\eta)\left(\theta_{J,p_0}+\frac{\theta_{J,(p_0-1)J}\theta_{(p_0-1)J,p_0}}{\phi(p_0J)}\right)^2,
\end{equation}
and
$$p_0J\log p=o_p\left(\frac{n}{J\Phi(1)}\left(\theta_{J,p_0}+\frac{\theta_{J,(p_0-1)J}\theta_{(p_0-1)J,p_0}}{\phi(p_0J)}\right)^2\beta_{\min}^2\right).$$
Then each step of the greedy FR will incorporate at least one relevant predictor and thus all the relevant predictors will be included in at most $p_0$ steps.
\end{thm}

\begin{rmk}
The expressions of our assumptions can be simplified under restricted isometry constant $\delta_s$ which is defined as the smallest quantity such that
$$ \left\{\frac{\vx^T\vX^T\vX\vx}{n\|\vx\|^2}:\|\vx\|_0\le s,\vx\neq\vnull\right\}\subseteq[1-\delta_s,1+\delta_s].$$
For example, following from the fact that $\theta_{s_1,s_2}\le \delta_{s_1+s_2}$ (Lemma 1.1 of \cite{candestao05}), condition (\ref{eqn:thm2cond10}) is implied by
\begin{equation}\label{eqn:thm2condsim}
\frac{  (\delta_{p_0+J}(1+\delta_{p_0J})+\delta_{p_0J}^2)^2}{(1-\delta_{p_0J})^5}\le \frac{J}{p_0(1+\eta)(1+\delta_1)}.
\end{equation}
\end{rmk}
\begin{rmk}
Intuitively, since each step of the greedy FR includes more additional covariates, the probability that a relevant covariate is incorporated is higher than using the original FR. Mathematically, it is unclear whether (\ref{eqn:thm2cond10}) or (\ref{eqn:thm2condsim}) represent a less stringent assumption for larger values of $J$, as both sides of the equation are generally increasing with $J$. 
\end{rmk}

In practice, one needs to select a model along the solution path. As in \cite{wang07a} and \cite{chenchen08}, we use the BIC-type criterion defined as 
$$BIC(k)=n\,\log(\|\vQ_{\calM^{(k)}}\vy\|^2)+ (kJ)\,\log n.$$
Then we choose the final model as the one which minimizes $BIC(k)$ among $k=0,1,\ldots,[n/J]$. The following theorem shows the screening consistency property when we use
the BIC stopping criterion in the greedy forward regression.
\begin{thm}\label{thm:3}
Under the same conditions as in Theorem \ref{thm:1}, and the assumptions that 
$$J=o(n/\log n) \quad \mbox{and} \quad J=o(\frac{\phi^3(p_0K_0J)n^2\beta_{\min}^2}{2\Phi(J)\Phi(1)\|\vy\|^2\log n}),$$
we have
$$P(\calT\subseteq \calM^{(\hat k)})\rightarrow 1,$$
where $\displaystyle{\hat k=\argmin_{0\le k\le [n/J]} BIC(k)}$.
\end{thm}

\section{Simulation Results}\label{sec:sim}
In this section, we perform Monte Carlo simulations to evaluate finite sample performance of the proposed greedy  forward regression (GFR) algorithm for ultra-high dimensional variable screening. We consider the following three simulation examples.

{\it Example 1} In this example, the components of $\vX=(X_1,\ldots,X_p)^T$ are generated from a multivariate normal distribution
$N(0,\Sigma)$, and $\Sigma$ is a block diagonal covariance matrix with $2\times 2$ blocks
$
\left(
\begin{array}{cc}
1& -0.4\\
-0.4 & 1
\end{array}
\right)
$.
The size  of the true model is chosen to be  $p_0=8$ with $\vbeta=(2,3,2,3,2,3,2,3,0,\ldots,0)^{T}$.

{\it Example 2} (Autoregressive correlation). For this simulation example, 
$\vX$ is a $p$-dimensional multivariate normal random vector with mean zero and
covariance matrix $(\sigma_{ij})$ with $\sigma_{ij}=0.5^{|i-j|}$ for $1\le i,j\le p$. The 1st,  4th and
7th components of $\vbeta$ are 3, 1.5 and 2, respectively, and the other elements of $\vbeta$ are fixed to be zero.

{\it Example 3}.  Consider  Example III in Section 4.2.3 of
Fan and Lv (2008) with
\begin{equation}\label{example 3}
Y=5X_1+5X_2+5X_3-15\sqrt{0.5}X_4+X_5+\epsilon,
\end{equation}
where $(X_1,X_2,X_3,X_6,\ldots,X_p)^{T}$ are generated from a multivariate normal distribution
$N(0,\Sigma)$ with entries of $\Sigma=(\sigma_{ij})_{(p-2)\times (p-2)}$
being $\sigma_{ii}=1,i=1,\ldots,p-2$ and $\sigma_{ij}=0.6, i\neq j$, and  $X_4\sim N(0,1)$ has correlation coefficient $\sqrt{0.5}$
with all the other $p-1$ variables whereas  $X_5\sim N(0,1)$ is
uncorrelated with all the other $p-1$ variables. In this example the true variable $X_5$ has an even weaker marginal correlation with $y$ than the irrelevant variables 
$X_6,\ldots,X_p$ do.

In all the above three examples, the noise $\epsilon$ is generated from a normal distribution with mean 0 and variance $\sigma^2$, and the variance $\sigma^2$ is selected so that the $R^2={\rm Var}\{\vX^{T}\vbeta\}/{\rm Var}(y)$ is approximately 50\%, 70\% or 90\%.  We considered sample size $n=150$ and three different predictor dimensions ($p=500, 1000$ or $2000$). For each case, we repeated the experiment 200 times. For GFR, we used $J=1, 2$ or 4. Obviously the GFR method reduces to the FR method proposed by \cite{wang09fr} when $J=1$.

Let $\hat{\vbeta}_{(k)}=(\hat{\beta}_{1(k)},\ldots,\hat{\beta}_{p(k)})^{T}\in \mathbb{R}^p$ denote the estimator obtained in the $k$th simulation replication (using some stopping criterion). The selected model is taken as $\widehat{\mathcal{M}}_{(k)}=\{j: |\hat{\beta}_{j(k)}|>0, j=1,\ldots,p\}$. We use the following performance measures to evaluate the methods:
{(1) The average number of false positives (AFP); (2) The average number of false negatives (AFN);}
(3) The average model size (AMS)  $200^{-1}\sum_k|\widehat{\mathcal{M}}_{(k)}|$; (4) The coverage probability (CP) $200^{-1}\sum_k I(\calT\subseteq \widehat{\calM}_{(k)})$.

The simulation study was carried out using MATLAB on a desktop computer with 3.20GHz CPU and 4GB RAM and the results  under three scenarios are reported in Tables \ref{tab:ex1.1}--\ref{tab:ex3.3}. 

\textit{Scenario (i)} We ran the GFR procedure exactly $p_{0}$ iterations, where $p_{0}$ is the true number of nonzero coefficients of $\vbeta$ in the true model,
and compute the average computational time when running stops (Time, in seconds) and CP. This is mainly to illustrate our Theorem \ref{thm:2} to see whether all the true nonzero coefficients can be identified in exactly $p_0$ steps. The results are reported in Tables \ref{tab:ex1.1}, \ref{tab:ex2.1} and \ref{tab:ex3.1}, for the three examples, respectively. We can see from these tables that when increasing from $J=1$ to either $J=2$ or $J=4$, most of the time there is a significant increase in CP  while only small additional cost in computational time is needed. 

\textit{Scenario (ii)} We ran the GFR till the end (incorporating close to $n$ variables in the model) and also recorded the time point when all relevant variables are incorporated into the estimated model (this time point is taken to be the time when running stops if not all relevant variables are incorporated when running stops). We computed the average computational time when running stops (time1), the average computational time when all nonzero coefficient are identified (time2), the average number of iterations (iter) and the average model size (AMS) when all nonzero coefficients are identified, and CP when running stops. The results are reported in Tables \ref{tab:ex1.2}, \ref{tab:ex2.2} and \ref{tab:ex3.2}. Compared to the case $J=1$, the CP values when running stops are often, although only slightly, larger when $J=2$ or 4. Note this is achieved with much shorter computational time (time1 reported in these tables). Also, the time to the point when all the relevant covariates are incorporated is in general shorter for larger values of $J$.

\textit{Scenario (iii)} Finally, we computed CP, AFP, AFN, AMS for the model selected by BIC,
 and the average computational time when running till BIC achieves its minimum value (time3). Note the AMS reported here is for the model selected based on BIC while in scenario (ii) the AMS is based on the model when all relevant covariates are incorporated.  We also included SIS and ISIS for comparison.  Following \cite{fan2008sure} and \cite{wang09fr},
the size of the SIS model was fixed to be $[n/\log n]$, and for
the ISIS method, a total of $[\log n-1]$ ISIS steps were conducted and $[n/ \log n]$ variables were selected in each step. 
The results are reported in Tables \ref{tab:ex1.3}, \ref{tab:ex2.3} and \ref{tab:ex3.3}.
The computational time (Time3) of GFR decreases as $J$ increases. In terms of the criteria  CP, AFP, AFN and AMS, the performances of GFR using different values of $J$ are similar. One exception is example 3 (Table \ref{tab:ex3.3}) where the CP for $J=4$ is low compared to $J=1,2$ when the signal is strong ($R^2=90\%$). However, by Scenario (ii), when running stops the CP for $J=4$ is satisfactory. This suggests that the problem resides in the fact that the BIC criterion stops the procedure too early. Finding a better criterion than BIC is thus an important problem, but very challenging one at the same time, which is outside the scope of the current paper. In Examples 1 and 3, SIS often has lower CP than GFR and FR do, but has higher CP in Example 2. However, note that the AMS (and AFP) of SIS is much larger than that of GFR. CP for ISIS is large for Examples 1 and 2, which is however achieved with even larger AMS and AFP.

\section{Applications to Real Datasets}\label{sec:app}

We applied GFR to two real data examples and compare it with FR, SIS and ISIS. First we considered the breast cancer dataset reported by \cite{vant2002}, which consists of 
expression levels for 24481 gene probes and seven clinical risk factors (age, tumor size, histological grade, angioinvasion, lymphocytic infiltration, estrogen receptor, and progesterone receptor status) for 97 lymph node-negative breast cancer patients 55 years old or younger. Among
the 97 patients, 46 developed distant metastases within 5 years and the other 51 remained metastases free for more than 5 years. \cite{yulima2012} proposed a ROC based approach to rank importance of the genes in predicting distant metastases after adjusting for the clinical risk factors. In their analysis, genes with severe missingness were removed,
and the other 24188 genes remained. The gene expression data were normalized such that all the variables have sample mean 0 and standard deviation 1.




Using their ranking methods \cite{yulima2012} found that, among the  24188 genes, gene 271 is the top one related to distant metastases within 5 years. Thus it is interesting to find genes that are related to gene 271.
Genes identified by FR, and the proposed GFR method with $J=2$ or $4$ are listed in Table \ref{tab:271gene}.
In addition, SIS found 21 genes which include all the genes identified by the GFR methods except gene 5342, and ISIS found 63 genes which include all the genes identified by GFR methods.
Then we compare the prediction mean squared errors (PMSE) of these different methods. For this purpose we randomly selected 90 observations as the training set and used the rest 7 observations for testing purpose. This procedure was repeated 20 times. The average of PMSEs over the 20 repetitions are reported in Table \ref{tab:271pmse}. From the table  we observe that GFR with $J=4$ has the smallest prediction error. Furthermore, note that ISIS has larger PMSE than SIS does, possibly because it includes more unimportant variables in the model.


Next we applied the proposed screening methods to a dataset arising from a microarray study in which expression quantitative trait locus (eQTL) mapping in laboratory rats was used to investigate gene regulation in the mammalian eye
and to identify genetic variation relevant to human eye disease \citep{scheetz06}. This dataset contains expressions of 31042 probe sets on 120 rats. Our goal is to find probes that are related to that of gene TRIM32, which has been found to cause Bardet-Biedl syndrome. The probe from TRIM32, 1389163\_at, is thus used as the response variable.
Similar to \cite{huangma08ss}, 3000 probes with the largest variances in expression values were used as covariates in our analysis.

Probes found by FR, and the GFR method with $J=2$ or $4$ are listed in Table \ref{tab:ratgene}.
SIS found 25 probes which include all those identified by the GFR methods except 1392692\_at and 1378099\_at, and ISIS found 75 probes which include all those identified by GFR methods except 1378099\_at.
To compare the prediction mean squared errors of these different methods, we randomly split the data into a training set of size 80 and a testing set of size 40. This procedure was repeated 1000 times. The average PMSEs of the different methods over the 1000 repetitions are reported in Table \ref{tab:ratpmse}. Again, we can see that GFR with $J=4$ has the smallest prediction error whereas the prediction error of ISIS is significantly larger than that of the other methods.


\section{Conclusions and Future Studies}\label{sec:con}

We propose GFR, a modification of the FR for sure screening of variables in sparse ultra-high dimensional linear regression, and show its theoretical and numerical advantages over the original FR. The main message is that GFR is faster to compute and its performance is comparable in terms of CP, AFP and AFN. Besides, when the signal is weaker ($R^2$ is lower), the simulation results suggest that the GFR tends to pick up more true variables than the FR does (yielding smaller AFN) at the expense of slightly larger AFP only. In addition, GFR appears to cope with correlation between the covariates better than FR does. The theoretical insights into these phenomena deserve future study. As mentioned in Section \ref{sec:sim}, it remains an important and challenging problem to construct alternative stopping criteria in order to improve the performance in terms of CP and AFN. FR is  known to be in general better than SIS or ISIS in yield a parsimonious model. The real data examples presented in Section \ref{sec:app} indicate that the GFR preserves this property, and even improves on FR in terms of prediction error. The GFR approach may be extended to other parametric models, such as generalized linear regression, and even to semiparametric models such as varying coefficient and semivarying coefficient models. Such extensions are non-trivial, however, and require further study. 

\section{Technical proofs}\label{sec:prof}

\textbf{Proof of Theorem \ref{thm:1}.} We start by assuming that no relevant predictors are contained in $\calN^{(k+1)}$. 
Let $\vX_{(k+1)}=\vQ_{\calM^{(k)}}\vX_{\calN^{(k+1)}}=\vX_{\calN^{(k+1)}}-\vX_{\calM^{(k)}}(\vX_{\calM^{(k)}}^T\vX_{\calM^{(k)}})^{-1}\vX_{\calM^{(k)}}^T\vX_{\calN^{(k+1)}}$ be the projection of $\vX_{\calN^{(k+1)}}$ onto the space orthgonal to $span\{\calM^{(k)}\}$. We have
\begin{eqnarray}\label{eqn:decompP}
\vP_{\calM^{(k+1)}}
&=&\vP_{\calM^{(k)}\cup\calN^{(k+1)}}\nonumber\\
&=&\vP_{\calM^{(k)}}+\vX_{(k+1)}(\vX_{(k+1)}^T\vX_{(k+1)})^{-1}\vX_{(k+1)}^T\nonumber\\
&=&\vP_{\calM^{(k)}}+\vQ_{\calM^{(k)}}\vX_{\calN^{(k+1)}}(\vX_{\calN^{(k+1)}}^T\vQ_{\calM^{(k)}}\vX_{\calN^{(k+1)}})^{-1}\vX_{\calN^{(k+1)}}^T\vQ_{\calM^{(k)}},
\end{eqnarray}
where in the second equality above  we used that columns of $\vX_{\calM^{(k)}}$ and columns of $\vX_{(k+1)}$ are orthogonal.
Using (\ref{eqn:decompP}), the change of SSR in the $k$-th step is
\begin{eqnarray*}
\|\vQ_{\calM^{(k)}}\vy\|^2-\|\vQ_{\calM^{(k+1)}}\vy\|^2
&=&\vy^T(\vQ_{\calM^{(k)}}-\vQ_{\calM^{(k+1)}})\vy\\
&=&\vy^T(\vP_{\calM^{(k+1)}}-\vP_{\calM^{(k)}})\vy\\
&=&\|(\vX_{\calN^{(k+1)}}^T\vQ_{\calM^{(k)}}\vX_{\calN^{(k+1)}})^{-1/2}\vX_{\calN^{(k+1)}}^T\vQ_{\calM^{(k)}}\vy\|^2\\
&\ge&\frac{1}{n\Phi(J)}\|\vX_{\calN^{(k+1)}}^T\vQ_{\calM^{(k)}}\vy\|^2\\
&=&\frac{1}{n\Phi(J)}\sum_{j\in \calN^{(k+1)}}|\vX_j^T\vQ_{\calM^{(k)}}\vy|^2.
\end{eqnarray*}

For $j\in \calN^{(k+1)}$, we have
\begin{eqnarray*}
|\vX_j^T\vQ_{\calM^{(k)}}\vy|^2
&\ge& {n\phi(kJ+1)}|(\vX_j^T\vQ_{\calM^{(k)}}\vX_j)^{-1/2}\vX_j^T\vQ_{\calM^{(k)}}\vy|^2\\
&\ge& {n\phi(kJ+1)}\max_{j\in\calT\backslash\calM^{(k)}}|(\vX_j^T\vQ_{\calM^{(k)}}\vX_j)^{-1/2}\vX_j^T\vQ_{\calM^{(k)}}\vy|^2\\
&\ge& \frac{\phi(kJ+1)}{\Phi(1)}\max_{j\in\calT\backslash\calM^{(k)}}|\vX_j^T\vQ_{\calM^{(k)}}\vy|^2,\\
\end{eqnarray*}
where in the first inequality we used Lemma \ref{lem:eigen} and in the second inequality we used that $\calN^{(k+1)}$ contains the indices with the $J$ largest values of 
$$\|\vP_{\calM^{(k)}\cup\{j\}}\vy\|^2=\|\vP_{\calM^{(k)}}\vy\|^2+
|(\vX_j^T\vQ_{\calM^{(k)}}\vX_j)^{-1/2}\vX_j^T\vQ_{\calM^{(k)}}\vy|^2$$ 
(the above equality follows from the same arguments as for (\ref{eqn:decompP})) among all $j\in \calF\backslash\calM^{(k)}$ and that no relevant covariate is contained in $\calN^{(k+1)}$.
Thus we have
\begin{eqnarray}\label{eqn:qmq1}
\lefteqn{\|\vQ_{\calM^{(k)}}\vy\|^2-\|\vQ_{\calM^{(k+1)}}\vy\|^2}\nonumber\\
&\ge& \frac{\phi(kJ+1)}{n\Phi(J)\Phi(1)}\cdot J \cdot \max_{j\in\calT\backslash\calM^{(k)}}|\vX_j^T\vQ_{\calM^{(k)}}\vy|^2\nonumber\\
&\ge&\frac{\phi(kJ+1)}{n\Phi(J)\Phi(1)}\cdot J \cdot \left(\max_{j\in\calT\backslash\calM^{(k)}}|\vX_j^T\vQ_{\calM^{(k)}}\vX_{\calT\backslash\calM^{(k)}}\vbeta_{\calT\backslash\calM^{(k)}}|-  \max_{j\in\calT\backslash\calM^{(k)}}|\vX_j^T\vQ_{\calM^{(k)}}\vepsilon|\right)^2
\end{eqnarray}
where the last inequality follows from $\vQ_{\calM^{(k)}}\vX_{\calM^{(k)}}=\vnull$.
Furthermore, let $t_k$ be the number of truly relevant covariates in $\calM^{(k)}$, we have
\begin{eqnarray}\label{eqn:qmq2}
&& \max_{j\in\calT\backslash\calM^{(k)}}|\vX_j^T\vQ_{\calM^{(k)}}\vX_{\calT\backslash\calM^{(k)}}\vbeta_{\calT\backslash\calM^{(k)}}|^2\nonumber\\
&\ge&\frac{1}{p_0-t_k}\|\vX_{\calT\backslash\calM^{(k)}}^T\vQ_{\calM^{(k)}}\vX_{\calT\backslash\calM^{(k)}}\vbeta_{\calT\backslash\calM^{(k)}}\|^2\nonumber\\
&\ge&\frac{n^2\phi^2(p_0-t_k+kJ)}{p_0-t_k}\|\vbeta_{\calT\backslash\calM^{(k)}}\|^2\nonumber\\
&\ge&n^2\phi^2(p_0-t_k+kJ)\beta_{\min}^2,
\end{eqnarray}
using Lemma \ref{lem:eigen}.

Under the subgaussian assumption of the noise, and noticing that $\|\vX_j^T\vQ_{\calM^{(k)}}\|^2\le \|\vX_j\|^2\le n\Phi(1)$, we have
$$P(|\vX_j^T\vQ_{\calM^{(k)}}\vepsilon|>t)\le c_1\exp\{-c_2t^2/(n\Phi(1))\},$$
and by the union bound
\begin{equation}\label{eqn:qmq3}
P(\sup_{j\in \calF, |\calM|\le M}|\vX_j^T\vQ_{\calM}\vepsilon|>t)=O(\exp\{-c_2t^2/(n\Phi(1))-M\log p\}).
\end{equation}
If (as we have assumed)
\begin{equation}\label{eqn:cond2}
p_0K_0J\log(p)=o_p(\frac{n\phi^2(p_0K_0J)\beta_{\min}^2}{\Phi(1)}),
\end{equation}
the second term  in (\ref{eqn:qmq1}) is dominated by the first term in (\ref{eqn:qmq1}).
Then for $k=1,\ldots,p_0K_0$, we have by (\ref{eqn:cond2}) above and (\ref{eqn:qmq1})-(\ref{eqn:qmq3}) that
$$\|\vQ_{\calM^{(k)}}\vy\|^2-\|\vQ_{\calM^{(k+1)}}\vy\|^2\ge \frac{\phi(p_0K_0J)J}{n\Phi(J)\Phi(1)}\frac{n^2\phi^2(p_0K_0J)\beta_{\min}^2}{2},$$
if step $k$ does not incorporate any relevant covariate.

Since $K_0$ is such that
\begin{equation*}
K_0> \frac{\|\vy\|^2}{ \frac{\phi(p_0K_0J)J}{n\Phi(J)\Phi(1)}\frac{n^2\phi^2(p_0K_0J)\beta_{\min}^2}{2}},
\end{equation*}
we see that within every $K_0$ steps there is at least one relevant covariate included. Thus after at most $p_0K_0$ steps all the relevant covariates are included (with the total number of covariates included at most $p_0K_0J$). \hfill $\Box$
\begin{lem}\label{lem:eigen}
For two models $\calM_1,\calM_2$, with $\calM_1\cap\calM_2=\emptyset$ and $|\calM_1\cup\calM_2|=s$, we have $\inf_{\|\vu\|=1}\vu^T\vX_{\calM_1}^T\vQ_{\calM_2}\vX_{\calM_1}\vu\ge n\phi(s)$.
\end{lem}
\textbf{Proof.} We write
$$\vQ_{\calM_2}\vX_{\calM_1}=\vX_{\calM_1}-\vX_{\calM_2}(\vX_{\calM_2}^T\vX_{\calM_2})^{-1}\vX_{\calM_2}^T\vX_{\calM_1}=(\vX_{\calM_1},\vX_{\calM_2})\left(\begin{array}{c}
					\vI \\
					-(\vX_{\calM_2}^T \vX_{\calM_2})^{-1}\vX_{\calM_2}^T \vX_{\calM_1}
							\end{array}\right). $$
Then
\bse
&&\vu^T\vX_{\calM_1}^T\vQ_{\calM_2}\vX_{\calM_1}\vu\\
&=&\vu^T[\vI, -\vX_{\calM_1}^T\vX_{\calM_2}(\vX_{\calM_2}^T \vX_{\calM_2})^{-1}] (\vX_{\calM_1\cup\calM_2}^T\vX_{\calM_1\cup\calM_2})  [\vI, -\vX_{\calM_1}^T\vX_{\calM_2}(\vX_{\calM_2}^T \vX_{\calM_2})^{-1}]^T\vu.
\ese
Since $\|\vu^T[\vI, -\vX_{\calM_1}^T\vX_{\calM_2}(\vX_{\calM_2}^T \vX_{\calM_2})^{-1}]\|\ge \|\vu\|=1$, and the smallest eigenvalue of \\ 
$\vX_{\calM_1\cup\calM_2}^T\vX_{\calM_1\cup\calM_2}$ is bounded below by $n\phi(s)$, thus we see  the claim is true. \hfill $\Box$\\

\noindent\textbf{Proof of Theorem \ref{thm:2}. }  Suppose each of the first $k$ steps identifies at least one relevant covariate. Assume $t_k$, the number of relevant covariates in $\calM^{(k)}$, is still less than $p_0$. Consider step $k+1$ of the algorithm. Let $R_j^{(k)}=\|\vP_{\calM^{(k)}\cup\{j\}}\vy\|^2$. To show that the $k$-th step also identifies at least one relevant covariate, we only need to show that $\max_{j\in\calT\backslash\calM^{(k)}}R_j^{(k)}$ is larger than the $J$-th largest value of $\{R_j^{(k)}: j\in \calT^c\backslash \calM^{(k)}\}$.

Similar to (\ref{eqn:decompP}), we have
\begin{equation}\label{eqn:decompP2}
\vP_{\calM^{(k)}\cup\{j\}}=\vP_{\calM^{(k)}}+\vQ_{\calM^{(k)}}\vX_j(\vX_j^T\vQ_{\calM^{(k)}}\vX_j)^{-1}\vX_j^T\vQ_{\calM^{(k)}},
\end{equation}
which gives
\begin{eqnarray*}
R_j^{(k)}
&=&\|\vP_{\calM^{(k)}}\vy\|^2+\left\|\frac{\vQ_{\calM^{(k)}}\vX_j\vX_j^T\vQ_{\calM^{(k)}}}{\|\vQ_{\calM^{(k)}}\vX_j\|^2}\vy\right\|^2\\
&=&\|\vP_{\calM^{(k)}}\vy\|^2+\frac{|\vX_j^T\vQ_{\calM^{(k)}}\vy|^2}{\|\vQ_{\calM^{(k)}}\vX_j\|^2}.
\end{eqnarray*}
Thus
\begin{eqnarray*}
&&\max_{j\in\calT\backslash\calM^{(k)}}R_j^{(k)}\\
&\ge&\|\vP_{\calM^{(k)}}\vy\|^2+\frac{1}{n\Phi(1)}\max_{j\in\calT\backslash\calM^{(k)}}|\vX_j^T\vQ_{\calM^{(k)}}\vy|^2\\
&\ge&\|\vP_{\calM^{(k)}}\vy\|^2+\frac{1}{n\Phi(1)}\left(\max_{j\in\calT\backslash\calM^{(k)}}|\vX_j^T\vQ_{\calM^{(k)}}\vX_{\calT\backslash\calM^{(k)}}\vbeta_{\calT\backslash\calM^{(k)}} |- \max_{j\in\calT\backslash\calM^{(k)}}|\vX_j^T\vQ_{\calM^{(k)}}\vepsilon |\right)^2,
\end{eqnarray*}
Using (\ref{eqn:qmq2}), we have
$$\max_{j\in\calT\backslash\calM^{(k)}}|\vX_j^T\vQ_{\calM^{(k)}}\vX_{\calT\backslash\calM^{(k)}}\vbeta_{\calT\backslash\calM^{(k)}}|\ge \frac{n\phi(p_0-t_k+kJ)}{\sqrt{p_0-t_k}}\|\vbeta_{\calT\backslash\calM^{(k)}}\|.$$

On the other hand, letting $\calJ$ be the set of indices of the $J$ largest values of $\{R_j^{(k)}: j\in \calT^c\backslash \calM^{(k)}\}$, then the $J$-th largest value of $\{R_j^{(k)}: j\in \calT^c\backslash \calM^{(k)}\}$ is bounded above by
\begin{eqnarray*}
&&\|\vP_{\calM^{(k)}}\vy\|^2+\frac{1}{J}\sum_{j\in\calJ}\frac{|\vX_j^T\vQ_{\calM^{(k)}}\vy|^2}{\|\vQ_{\calM^{(k)}}\vX_j\|^2}\\
&\le &\|\vP_{\calM^{(k)}}\vy\|^2+\frac{1}{nJ\phi(kJ+1)}\sum_{j\in\calJ}|\vX_j^T\vQ_{\calM^{(k)}}\vy|^2\\
&\le &\|\vP_{\calM^{(k)}}\vy\|^2+\frac{1+\eta}{nJ\phi(kJ+1)}\sum_{j\in\calJ}\left(|\vX_j^T\vQ_{\calM^{(k)}} \vX_{\calT\backslash\calM^{(k)}}\vbeta_{\calT\backslash\calM^{(k)}}   |^2+\frac{1}{\eta}|\vX_j^T\vQ_{\calM^{(k)}} \vepsilon|^2\right),
\end{eqnarray*}
for any $\eta>0$.
Furthermore,
\begin{eqnarray*}
&&\sum_{j\in\calJ}|\vX_j^T\vQ_{\calM^{(k)}} \vX_{\calT\backslash\calM^{(k)}}\vbeta_{\calT\backslash\calM^{(k)}}|^2\\
&=&\|\vX_{\calJ}^T\vQ_{\calM^{(k)}} \vX_{\calT\backslash\calM^{(k)}}\vbeta_{\calT\backslash\calM^{(k)}}\|^2\\
&\le&(\|\vX_{\calJ}^T\vX_{\calT\backslash\calM^{(k)}}\vbeta_{\calT\backslash\calM^{(k)}}\|+\|\vX_{\calJ}^T\vX_{\calM^{(k)}}(\vX_{\calM^{(k)}}^T\vX_{\calM^{(k)}})^{-1}\vX_{\calM^{(k)}}^T \vX_{\calT\backslash\calM^{(k)}}\vbeta_{\calT\backslash\calM^{(k)}}\|)^2\\
&\le& (n\theta_{J,p_0-t_k}\|\vbeta_{\calT\backslash\calM^{(k)}}\|+\frac{n\theta_{J,kJ}\theta_{kJ,p_0-t_k}}{\phi(kJ)}\|\vbeta_{\calT\backslash\calM^{(k)}}\|)^2
\end{eqnarray*}
Thus, we have that the $J$-th largest value of $\{R_j^{(k)}: j\in \calT^c\backslash \calM^{(k)}\}$ is bounded above by
\begin{eqnarray*}
& \|\vP_{\calM^{(k)}}\vy\|^2+\frac{1+\eta}{nJ\phi(kJ+1)}(n\theta_{J,p_0-t_k}+\frac{n\theta_{J,kJ}\theta_{kJ,p_0-t_k}}{\phi(kJ)})^2\|\vbeta_{\calT\backslash\calM^{(k)}}\|^2\\
& +\frac{1+\eta}{n\eta\phi(kJ+1)}\max_{j\in\calT^c\backslash\calM^{(k)}}|\vX_j^T\vQ_{\calM^{(k)}} \vepsilon|^2.
\end{eqnarray*}
Thus if
\begin{equation}\label{eqn:thm2cond1}
\frac{n^2\phi^2(p_0-t_k+kJ)}{n\Phi(1)(p_0-t_k)}\ge\frac{1+\eta}{nJ\phi(kJ+1)}\left(n\theta_{J,p_0-t_k}+\frac{n\theta_{J,kJ}\theta_{kJ,p_0-t_k}}{\phi(kJ)}\right)^2,
\end{equation}
for some constant $\eta>0$,
and
\begin{equation}\label{eqn:thm2cond2}
\max_{j\in\calF}|\vX_j^T\vQ_{\calM^{(k)}} \vepsilon|^2=o_p\left(\frac{1}{J}\left(n\theta_{J,p_0-t_k}+\frac{n\theta_{J,kJ}\theta_{kJ,p_0-t_k}}{\phi(kJ)}\right)^2\|\beta_{\calT\backslash\calM^{(k)}}\|^2\right),
\end{equation}
then at least one relevant predictor will be selected  in step $k+1$. 
Noting $t_k\ge k$ and $p_0-t_k+kJ\le p_0J$, (\ref{eqn:thm2cond1}) is implied by
$$
\frac{\phi^3(p_0J)J}{\Phi(1)p_0}\ge (1+\eta)\left(\theta_{J,p_0}+\frac{\theta_{J,(p_0-1)J}\theta_{(p_0-1)J,p_0}}{\phi(p_0J)}\right)^2.
$$
In addition, by (\ref{eqn:qmq3}), (\ref{eqn:thm2cond2}) is implied by
$$p_0J\log p=o_p\left(\frac{n}{J\Phi(1)}\left(\theta_{J,p_0}+\frac{\theta_{J,(p_0-1)J}\theta_{(p_0-1)J,p_0}}{\phi(p_0J)}\right)^2\beta_{\min}^2\right).$$
\hfill $\Box$

\noindent\textbf{Proof of Theorem \ref{thm:3}. } By Theorem \ref{thm:1}, we know that $\calT\subseteq \calM^{p_0K_0}$ with probability approaching 1. We only need to show that
$$P\Big(  \min_{\calT\backslash \calM^{(k)}\neq\emptyset, k\le p_0K_0} BIC(k)-BIC(k+1)>0\Big)\rightarrow 1.$$
We have shown in the proof of Theorem \ref{thm:1} that if $\calN^{(k+1)}\cap\calT=\emptyset$,
$$\|\vQ_{\calM^{(k)}}\vy\|^2-\|\vQ_{\calM^{(k+1)}}\vy\|^2\ge \frac{\phi(p_0K_0J)J}{n\Phi(J)\Phi(1)}\frac{n^2\phi^2(p_0K_0J)\beta_{\min}^2}{2}.$$
On the other hand, if $\calN^{(k+1)}\cap\calT\neq\emptyset$, using almost the same arguments, we have with probability approaching 1,
\begin{eqnarray*}
\|\vQ_{\calM^{(k)}}\vy\|^2-\|\vQ_{\calM^{(k+1)}}\vy\|^2
&\ge&\frac{1}{n\Phi(J)}\sum_{j\in \calN^{(k+1)}}|\vX_j^T\vQ_{\calM^{(k)}}\vy|^2\\
&\ge&\frac{1}{n\Phi(J)}\max_{j\in \calN^{(k+1)}}|\vX_j^T\vQ_{\calM^{(k)}}\vy|^2\\
&\ge& \frac{\phi(kJ+1)}{n\Phi(J)\Phi(1)} \cdot \max_{j\in\calT\backslash\calM^{(k)}}|\vX_j^T\vQ_{\calM^{(k)}}\vy|^2\\
&\ge&\frac{\phi(p_0K_0J)}{n\Phi(J)\Phi(1)}\frac{n^2\phi^2(p_0K_0J)\beta_{\min}^2}{2}.
\end{eqnarray*}
Note the only difference from the $\calN^{(k+1)}\cap\calT=\emptyset$ case is the removal of a factor of $J$ in the lower bound.
Then
\begin{eqnarray*}
BIC(k)-BIC(k+1)
&=&n\log(\|\vQ_{\calM^{(k)}}\vy\|^2)-n\log(\|\vQ_{\calM^{(k+1)}}\vy\|^2)-J\log(n)\\
&=&n\log\left(1+\frac{\|\vQ_{\calM^{(k)}}\vy\|^2-\|\vQ_{\calM^{(k+1)}}\vy\|^2}{\|\vQ_{\calM^{(k+1)}}\vy\|^2}\right)-J\log(n).
\end{eqnarray*}
Using the elementary inequality $\log(1+x)\ge \min\{\log 2,x/2\}$, the lower bound for $\|\vQ_{\calM^{(k)}}\vy\|^2-\|\vQ_{\calM^{(k+1)}}\vy\|^2$ above, and the fact $\|\vQ_{\calM^{(k+1)}}\vy\|^2\le \|\vy\|^2$, we get
\begin{eqnarray*}
BIC(k)-BIC(k+1)
&=&\min\left\{n\log 2, \frac{\phi^3(p_0K_0J)n^2\beta_{\min}^2}{2\Phi(J)\Phi(1)\|\vy\|^2}\right\} -J\log  n \,.
\end{eqnarray*}
Under our assumptions, the quantity on the RHS of the above equality is positive with probability approaching 1. Hence the proof is completed. \hfill $\Box$


\bibliographystyle{dcu}
\bibliography{papers}

\newpage
\clearpage

\begin{table}[htb]
 \begin{center}
\caption{\label{tab:ex1.1}  Simulation results of Example 1, scenario (i). }
\vspace{0.1in}
\begin{tabular}{ cccccccc }
\hline && \multicolumn{2}{c}{FR}& \multicolumn{2}{c}{GFR($J=2$)}&\multicolumn{2}{c}{GFR($J=4$)}\\
\cline{3-4}\cline{5-6}\cline{7-8}
 $p$ & $R^2$ & CP & Time (s) & CP& Time (s) & CP & Time (s)   \\
 \hline
 500 &      &0.9950  & 1.1399 & 1 &1.4808  &1&1.8915 \\
  1000   & $90\%$ & 0.9850 &2.3243   & 1 & 3.0221 & 1&3.8769\\
   2000  &    & 0.9950 &4.7588  & 1 & 6.1619 &1&7.9348 \\ \hline  
   500   &     & 0.9100   &1.1315   & 0.9850&1.4753  &0.9950&1.8946 \\
   1000  &  $70\%$ &0.8250   &2.3116  & 0.9700 &3.0089  &0.9500&3.8820 \\
  2000 &   & 0.6850 &4.7530  & 0.9250 & 6.1739 &0.9300&7.9421 \\ \hline
   500    &      & 0.1350 & 1.1316 &0.4350 &1.4797 & 0.4600&1.8943\\
  1000  &  $50\%$ &0.0600 & 2.3079 & 0.1600  & 3.0153 &0.2350&3.8792  \\
  2000  &  &0.0150&4.7561 &0.0250&6.1693 & 0.0350&7.9297\\
\hline
\end{tabular}%
 \end{center}
\end{table}

\clearpage
\begin{table}[htb]
 \begin{center}
 \caption{\label{tab:ex1.2}  Simulation results of Example 1, scenario (ii).}
\vspace{0.1in}
\begin{tabular}{ cccccccc }
\hline
 Method &$p$ &  $R^2$ & CP  & AMS & iter& Time1 (s) & Time2 (s) \\
 \hline
 FR&500 &  &1  &8.0100 &8.0100&32.9721  &1.1024\\
 & 1000   & $90\%$ & 1   &8.0200 &8.0200&68.9648 &2.2070 \\
 &  2000  &    & 1  &8.0100 & 8.0100& 141.7676&4.4289\\
 &  500   &   &0.9950  &8.3878 &8.3878&32.8481  &1.8205 \\
 &  1000  & $70\%$  & 1  &8.2400 &8.2400&69.1548 &2.3230 \\
 & 2000 &   & 0.9500  &  8.3469&  8.3469&141.4304& 7.4696 \\
 &  500    &   &0.5250  & 16.8889&16.8889& 33.0695 &17.3780 \\
 & 1000  &  $50\%$  &0.2100   &16.6000 &16.6000&69.5911 &57.0291 \\
 & 2000  &  &0.0600  & 24.6667&24.6667& 142.5579 &135.4758\\
 \\
GFR($J=2$) & 500 &  & 1 & 8.0600 & 4.0300 & 17.0107 &0.4833 \\
 & 1000   & $90\%$ &1  &8.1000  & 4.0500 & 35.3227 &0.9541 \\
 &  2000  &    & 1   &8.1200  & 4.0600&72.4462 &1.9353 \\
 &  500   &  &1   &8.4400  &4.2200&16.8715   & 0.5194 \\
 &  1000  &  $70\%$  & 1 &9.0100  &4.5050& 35.3962   &1.1790 \\
 & 2000 &   & 0.9350 &9.7778  &4.8889& 72.8293   &9.7143\\
 & 500    &   &0.5800  & 14.8334 &7.4167& 16.9897   &9.5254 \\
 & 1000  &  $50\%$  & 0.2250   & 21.4546 & 10.7273&35.7115 & 28.9731 \\
 & 2000  &  & 0.0800    & 17.5000 & 8.2500&72.4546 &67.5950\\
 \\
GFR($J=4$) &500 &  &1  & 8.9600 & 2.2400& 8.2202  & 0.3185 \\
 & 1000   & $90\%$ & 1   &9.4600  &2.3650& 17.3058 &0.6937\\
 &  2000  &    & 1  & 9.7800 &2.4450  &35.5461&1.5884  \\
 &  500   &   & 1   &11.5600  &2.8900 &8.2217 &0.4644 \\
 &  1000  &  $70\%$ & 1   &12.4252  &3.1063& 17.3298 &1.0364 \\
 & 2000 &   & 0.9200    &12.7112  &3.1778& 35.5907 & 5.4671 \\
 &  500    &   & 0.5450   &19.4544  &4.8636 &8.2735 &5.0755 \\
 & 1000  & $50\%$   & 0.2150    &21.6000  &5.4000&17.4886 &14.4169\\
 & 2000  &  & 0.0900   & 27.6200 &6.9050&35.9331 &35.0428\\
\hline
\end{tabular}%

\end{center}
\end{table}

\clearpage
\begin{table}[htb]
 \begin{center}
 \caption{\label{tab:ex1.3}  Simulation results of Example 1, scenario (iii). } 
\vspace{0.1in}
\resizebox*{!}{\dimexpr\textheight-2\baselineskip\relax}
{
\begin{tabular}{ cccccccc }
\hline
 Method &$p$ &  $R^2$ & CP & AFP & AFN & AMS &Time3 (s)\\
 \hline
 FR &500 &  & 1 & 0.1200& 0   &  8.1200 & 1.1145 \\
    &1000&    $90\%$   &  1  &0.0100 & 0   &  8.0100  &2.1929\\
    &2000&  &  1  &0.0200  &0  &       8.0200  &4.4733\\  
    &500 &  & 0.5750 & 0.0500  & 0.3500& 7.7000 &1.0480\\    
    &1000& $70\%$ &  0.4500 &0.3200 & 1.4300 & 6.8900  &2.0102\\
    &2000&  &  0.2550 &0.1500 & 2.6900 & 5.4600  &3.6044\\   
    &500 &  & 0 & 0.0200  & 5.1400 & 2.8800 &0.2783\\
    &1000& $50\%$ &  0  &0.0200  & 5.9000 & 2.1200  &0.3805\\
    &2000&  &  0  &0.0500  &6.3600 & 1.6900  &0.5836\\
\\
GFR($J=2$) &500 && 1 & 0.0400 & 0   & 8.0400 &  0.4675 \\
    &1000&  $90\%$  &  1  &0.0800 &   0   & 8.0800  & 0.9451 \\
    &2000&  &  1  &0.0800   &   0 & 8.0800 & 1.9091\\
    &500 &  & 0.8400 & 0.2300  &  0.4500    &  7.7800 &  0.4613\\
    &1000& $70\%$ &  0.5300  &0.3000   & 1.5800  & 6.7200   & 0.7660\\
    &2000&  &  0.3900 &0.5200   & 2.2200    & 6.3000 &1.4399 \\ 
    &500 &  & 0.0100 &  0.3500  &  5.8500  & 2.5000 &  0.1035 \\
    &1000& $50\%$ &  0  &0.4300   & 6.2500  & 2.1800  & 0.1704 \\
    &2000&  &  0  &0.5200   & 6.3800    & 2.1400  & 0.3296 \\
\\
GFR($J=4$)&500 &  & 1 & 1.2800   & 0  & 9.2800 &  0.3210\\
    &1000& $90\%$ &  1  &1.7600   & 0    & 9.7600 & 0.6962\\
    &2000&  &  1  &1.8400   & 0  & 9.8400 &  1.4139\\   
    &500 & & 0.7300 & 1.3400  &  1.4200   & 7.9200& 0.2482\\
    &1000& $70\%$  &0.3700 &  1.3000  &2.7400   & 6.5600  & 0.3740\\
    &2000&  &  0.2950 &1.3800   & 4.1400  & 5.2400  &  0.5153\\
    &500 & & 0.0100 & 1.4000  &  5.2800   & 4.1200& 0.0804 \\
    &1000& $50\%$  &  0  & 1.5600   & 5.5200   & 4.0400 &  0.1545 \\
    &2000&  &  0  &1.9800   & 5.9400    & 4.0400 & 0.3090\\
\\
    SIS &500 &  &0.0200  &  23.4500   & 2.4500   &     29.0000&0.0218 \\
    &1000& $90\%$ &  0.0050  & 24.0250  &  3.0250 &   29.0000 & 0.0786 \\
    &2000&  & 0  &24.2600  &  3.2600  &     29.0000& 0.2974\\  
    &500 &  &  0.0100   & 23.7900   & 2.7900   &     29.0000  &0.0219 \\
    &1000& $70\%$ & 0   & 24.3350 &   3.3350  &    29.0000& 0.0784 \\
    &2000&  & 0    & 24.6650   & 3.6650& 29.0000 &  0.3004 \\   
    &500 & & 0   & 24.3000   & 3.3000& 29.0000 &  0.0248 \\
    &1000&  $50\%$ &  0 & 24.8600  &  3.8600 &  29.0000 &0.0899 \\
    &2000&  &0  & 25.3300   & 4.3300&  29.0000 &0.3269 \\
\\
ISIS &500 &  &1   & 108.0000   &      0  &   116.0000&0.0530 \\
    &1000& $90\%$ &  1   & 108.0000   &      0&   116.0000 & 0.1321 \\
    &2000&  & 0.9850   & 108.0250   &   0.0250 &  116.0000& 0.4095 \\
    &500 &  & 0.8900     &  108.1150  &  0.1150& 116.0000  &0.0509 \\
    &1000& $70\%$ & 0.6700    & 108.4350  &  0.4350  & 116.0000 & 0.1301 \\
    &2000&  & 0.3250      &109.1250  &  1.1250  & 116.0000&  0.4103 \\  
    &500 & & 0.3950  & 108.8400   & 0.8400  & 116.0000  &  0.0519 \\
    &1000&  $50\%$ &  0.1000   & 109.9250   & 1.9250 & 116.0000 &0.1279 \\
    &2000&  & 0.0200  &111.0350   & 3.0350  &  116.0000   &0.3982 \\
\hline
\end{tabular}%
}

\end{center}
\end{table}

\clearpage
\begin{table}[htb]
 \begin{center}
\caption{\label{tab:ex2.1}  Simulation results of Example 2, scenario (i). }
\vspace{0.1in}
\begin{tabular}{ cccccccc }
\hline && \multicolumn{2}{c}{FR}& \multicolumn{2}{c}{GFR($J=2$)}&\multicolumn{2}{c}{GFR($J=4$)}\\
\cline{3-4}\cline{5-6}\cline{7-8}
 $p$ & $R^2$ & CP & Time (s) & CP& Time (s) & CP & Time (s)   \\
 \hline
 500 &      &1  & 0.2992 & 1 &0.3776  &1&0.5104 \\
  1000   & $90\%$ & 1 &0.7010   & 1 & 0.8487 & 1&1.1190\\
   2000  &    & 1 &4.3744  & 1 & 4.6684 &1&5.2827 \\ \hline
   500   &    & 0.9950   &0.3001   & 1&0.3791  &1&0.5083 \\
   1000  & $70\%$   &0.9850   &0.7114  & 0.9900 &0.8485  &1&1.1099 \\
  2000 &   & 0.9900 &4.3898  & 0.9850 & 4.6705 &0.9950&5.2033 \\ \hline
   500    &      & 0.7850 & 0.3014 &0.8400 &0.3787 & 0.9150&0.5199\\
  1000  & $50\%$  &0.7100 & 0.6925 & 0.8050  & 0.8560 &0.8800&1.1349 \\
  2000  &  &0.6950&4.3795 &0.7700&4.7253 & 0.8350&5.2680\\
\hline
\end{tabular}%
\end{center}
\end{table}

\clearpage
\begin{table}[htb]
 \begin{center}
 \caption{\label{tab:ex2.2}  Simulation results of Example 2, scenario (ii).}

\vspace{0.1in}
\begin{tabular}{ cccccccc }
\hline
 Method &$p$ &  $R^2$ & CP & AMS & iter& Time1 (s) & Time2 (s)  \\
 \hline
 FR&500 &  &1   &3.0000 &3.0000&32.8695&0.2599\\
 & 1000   & $90\%$ & 1   &3.0000 &3.0000&69.1921 &0.5219 \\
 &  2000  &    & 1  &3.0000 & 3.0000& 144.6914& 1.0535\\
 &  500   &   &1   &3.0050 &3.0050&32.6123   & 0.2606\\
 &  1000  & $70\%$  & 0.9950 &3.0101 &3.0101 &68.6967 &0.8658  \\
 & 2000 &   & 0.9950 &  3.0653&  3.0653&145.2301 &1.8179  \\
 &  500    &  &0.9500  & 3.8316&3.8316 & 32.9186 &  2.0148\\
 & 1000  &   $50\%$  &0.8500  &4.0000 &4.0000 &69.1772 &  10.9656 \\
 & 2000  &  &0.7200  & 3.5556&3.5556& 144.9864  &  40.2103\\
 \\
GFR($J=2$) & 500 &  & 1 & 4.0400 & 2.0200& 16.8159  &0.1746 \\
 & 1000   & $90\%$ &1   &4.0500  & 2.0250& 35.5480 &0.3521 \\
 &  2000  &    & 1   &4.0500  & 2.0250 &75.8053&0.7142 \\
 &  500   &   &1  &4.1500  &2.0750&16.8271   & 0.2146 \\
 &  1000  &  $70\%$ & 0.9950 &4.1106  &2.0553 & 35.4387  &0.4281 \\
 & 2000 &   & 0.9850  &4.0914  &2.0457 & 75.4025 &1.4179 \\
 &  500    &   &0.9250  & 6.3530 &3.1765& 16.9038   &  1.9379 \\
 & 1000  & $50\%$   & 0.8800   & 5.6316 & 2.8158 &35.6034& 5.0057 \\
 & 2000  &  & 0.8050   & 4.7000 & 2.3500&75.8922  &   17.2509 \\
 \\
GFR($J=4$) &500 &&1   & 4.5000 & 1.1250 & 8.2943&  0.0966 \\
 & 1000   &  $90\%$  & 1   &4.4600  &1.1150& 17.5272 &0.1903 \\
 &  2000  &    & 1  & 4.4000 &1.1000&38.8654  &0.3771  \\
 &  500   &  & 1   &4.9200  &1.2300 &8.2625  &  0.1213  \\
 &  1000  & $70\%$   & 1  &4.7600  &1.1900 & 17.3947   & 0.2149\\
 & 2000 &   & 0.9950   &4.9696 &1.2424& 36.0217   &   0.8492 \\
 &  500    &    & 0.9400  &5.7192  &1.4298&8.2867  &   0.7015 \\
 & 1000  &  $50\%$ & 0.9100   &5.1620  &1.2905 &17.8103  &  1.8052 \\
 & 2000  &  & 0.8850  & 5.5204 &1.3801 &39.0071&   6.5094\\
\hline
\end{tabular}%

\end{center}
\end{table}

\clearpage
\begin{table}[htb]
 \begin{center}
 \caption{\label{tab:ex2.3}  Simulation results of Example 2, scenario (iii). }
\vspace{0.1in}
\resizebox*{!}{\dimexpr\textheight-2\baselineskip\relax}
{
\begin{tabular}{ cccccccc }
\hline
 Method &$p$ &  $R^2$& CP & AFP & AFN & AMS & Time3 (s)\\
 \hline
 FR &500 &  &  1 &   0.0333       &  0   & 3.0350 & 0.2669\\
    &1000& $90\%$ &  1  & 0.0067    &     0   & 3.0050  &   0.5269 \\
    &2000&  &  1 &  0.0200       &  0     & 3.0200  & 1.0728 \\
    &500 & & 0.9950  & 0.0400 &   0.0050   &     3.0350  & 0.2652\\
    &1000& $70\%$  &  0.9700 &  0.0150   & 0.0350   &       2.9800  & 0.5190\\
    &2000&  & 0.9650  & 0.0250   & 0.0350   &        2.9900  & 1.0585\\
    &500 &  & 0.5000  &  0.0650  &  0.5400   & 2.5250 & 0.2158 \\
    &1000& $50\%$ & 0.3750  &0.0400  &  0.6750  &      2.3650  &  0.4006 \\
    &2000&  &  0.3800  & 0.0350  &  0.6600   &       2.3750 & 0.8133\\
\\
GFR($J=2$) &500 &  & 1.0000 & 1.0600    &     0    &  4.0600 &  0.1788 \\
    &1000& $90\%$ &  1.0000  & 1.0500    &     0     & 4.0500 & 0.3532\\
    &2000&  &  1.0000  &  1.0500      &   0    &       4.0500 &  0.7084 \\
    &500 &  & 0.8200  & 0.9700   & 0.1800   & 3.7900 & 0.1643 \\
    &1000& $70\%$ &  0.8150   & 0.9150   & 0.1850      &  3.7300 & 0.3194\\
    &2000&  &  0.7450  & 0.8400   & 0.2600  &    3.5800   & 0.6107\\
    &500 & &  0.2300  & 0.4800   & 0.9000     & 2.5800 & 0.1045  \\
    &1000&  $50\%$ &  0.2200  &0.3900   & 0.8700   &    2.5200  & 0.2026 \\
    &2000&  &  0.2000&  0.3500  &  0.8700   &     2.4800 & 0.4019 \\
\\
GFR($J=4$)&500 &  & 1.0000  & 1.5000     &    0   &  4.5000& 0.0969 \\
    &1000& $90\%$ &  0.9950 & 1.4450   & 0.0050    & 4.4400 &  0.1898 \\
    &2000&  &   1.0000 &  1.4000       &  0    &  4.4000 &0.3745 \\
    &500 & & 0.7950 &  1.2900   & 0.2100   &  4.0800 & 0.0793 \\
    &1000& $70\%$  &  0.8300  & 1.3700   & 0.1700  &        4.2000  &  0.1698\\
    &2000&  &   0.7950  & 1.2650   & 0.2050  &        4.0600  & 0.3158 \\
    &500 &  & 0.6300  & 1.3850    &0.3850  &      4.0000  & 0.0760 \\
    &1000& $50\%$ &  0.6800 & 1.3350   & 0.3350  &        4.0000 &   0.1532 \\
    &2000&  & 0.6250  &1.3850    &0.3850   &       4.0000 &  0.3057\\
\\

SIS &500 &  &1.0000    &  26.0000     &    0&  29.0000 &0.0216 \\
    &1000& $90\%$ &  1.0000   & 26.0000    &    0& 29.0000  & 0.0773 \\
    &2000&  & 1.0000  &  26.0000     &    0&29.0000  & 0.4183 \\
    &500 &  &  1.0000      &26.0000     &    0& 29.0000&0.0217 \\
    &1000& $70\%$ & 1.0000   & 26.0000 &0 &  29.0000 & 0.0775 \\
    &2000&  & 1.0000   &26.0000    &     0 &   29.0000  &  0.4191 \\
    &500 &  & 0.9950  &  26.0050    &0.0050 &  29.0000 &  0.0222 \\
    &1000& $50\%$ & 0.9950     & 26.0050   & 0.0050&29.0000 &0.0791 \\
    &2000&  & 0.9750  &  26.0250  &  0.0250&  29.0000 &0.4169 \\
\\
ISIS &500 &  &1.0000    & 113.0000     &    0   &  116.0000 &0.0538 \\
    &1000& $90\%$ &  1.0000     &113.0000    &     0& 116.0000   & 0.1347 \\
    &2000&  & 1.0000  & 113.0000      &  0  &  116.0000  & 0.5474 \\
    &500 &  &  1.0000    &   113.0000   &      0  &116.0000 &0.0504 \\
    &1000& $70\%$ & 1.0000    &  113.0000   &      0 &116.0000 & 0.1334 \\
    &2000&  & 1.0000    & 113.0000     &    0 &116.0000  &  0.5673 \\
    &500 &  & 0.9950  & 113.0050  &  0.0050 &  116.0000   &  0.0512 \\
    &1000& $50\%$ &0.9950  & 113.0050  &  0.0050  & 116.0000 &0.1266 \\
    &2000&  & 0.9800   & 113.0200    &0.0200&  116.0000 &0.5431 \\
\hline
\end{tabular}%
}

\end{center}
\end{table}

\clearpage
\begin{table}[htb]
 \begin{center}
\caption{\label{tab:ex3.1}  Simulation results of Example3, scenario (i). }
\vspace{0.1in}
\begin{tabular}{ cccccccc }
\hline && \multicolumn{2}{c}{FR}& \multicolumn{2}{c}{GFR($J=2$)}&\multicolumn{2}{c}{GFR($J=4$)}\\
\cline{3-4}\cline{5-6}\cline{7-8}
 $p$ & $R^2$ & CP & Time (s) & CP& Time (s) & CP & Time (s)   \\
 \hline
 500 &     &0.5000  & 0.5973 & 1 &0.7165  &1&0.9788 \\
  1000   & $90\%$  & 0.3050 &1.2458   & 0.9950 & 1.4566 & 0.9950&2.0121\\
   2000  &    & 0.2350 &2.6009  & 0.9950 & 3.0360 &0.9900&4.1696 \\  \hline
   500   &     & 0.0550   &0.6010   & 0.5850&0.7110  &0.705&0.9752 \\
   1000  &  $70\%$ &0.0100   &1.2486  & 0.4300 &1.4476  &0.4950&1.9991 \\
  2000 &   & 0.0200 &2.5964  & 0.3750 & 3.0176 &0.4400&4.1575 \\  \hline
   500    &      & 0.0050 & 0.5953 &0.2200 &0.7111 & 0.3450&0.9767\\
  1000  &  $50\%$ &0 & 1.2194 & 0.1550  & 1.4564 &0.2000&2.0005 \\
  2000  &  &0&2.5473 &0.0700&3.0350 & 0.1350&4.1425\\
\hline
\end{tabular}%
\end{center}
\end{table}

\clearpage
\begin{table}[htb]
 \begin{center}
 \caption{\label{tab:ex3.2}  Simulation results of Example 3, scenario (ii).}

\vspace{0.1in}
\begin{tabular}{ cccccccc }
\hline
 Method &$p$ &  $R^2$ & CP & AMS & iter& Time1 (s) & Time2 (s)  \\
 \hline
 FR&500 & &1   &5.5150 &5.5150&31.5283& 0.6470\\
 & 1000   & $90\%$  & 1   &5.8550 &5.8550 &66.4831&1.3978\\
 &  2000  &    & 1  &5.8995 & 5.8995& 136.5545&2.8402\\
 &  500   &   & 0.7250  &9.1379 &9.1379 &31.5440  &  12.1350\\
 &  1000  & $70\%$  & 0.5350 &11.9065 &11.9065 &66.5275  &  36.4303 \\
 & 2000 &   & 0.4300  &  8.8372&  8.8372  &136.7187  & 79.8748 \\
 &  500    &   & 0.2850 & 17.5263&17.5263 & 31.5433   & 23.3161 \\
 & 1000  & $50\%$   &0.1400   &17.2857 &17.2857 &66.5714 &  58.3595 \\
 & 2000  &  &0.0400 & 17.8750&17.8750  & 137.4194 &  131.7264\\
 \\
GFR($J=2$) & 500 & & 1  & 7.3000 &3.6500 & 16.1349 &  0.4118\\
 & 1000   &  $90\%$ &0.9950   &7.3970  & 3.6985  & 34.1177 & 0.8488 \\
 &  2000  &    & 0.9950   &7.5778  & 3.7889 &70.1851 &  2.4704 \\
 &  500   &   &0.7650  &12.1700  & 6.0850 &16.1258  &  5.2963 \\
 &  1000  & $70\%$  & 0.5450  &12.0918  &6.0459 & 34.0971  & 19.3409 \\
 & 2000 &   & 0.4450&10.2022  &5.1011  & 70.0645 &  43.2803 \\
 &  500    &   &0.3950  & 19.8228 &9.9114 & 16.1479  & 11.2915 \\
 & 1000  & $50\%$   & 0.2650   & 19.1320 &  9.5660 &34.0703  & 26.4529\\
 & 2000  &  & 0.1350   & 19.1112 & 9.5556 &70.1724   & 65.6157\\
 \\
GFR($J=4$) &500 & &1  & 11.2800 &2.8200  & 7.9370 &  0.4282 \\
 & 1000   & $90\%$  & 1  &11.5200  &2.8800 & 16.7251  &  0.8762 \\
 &  2000  &    & 0.9900 & 11.5352 &2.8838  &34.4208 &  2.1484 \\
 &  500   &   & 0.7950   &15.4968  &3.8742  &7.9275  & 2.4925 \\
 &  1000  & $70\%$  & 0.5800  &14.9312  &3.7328 & 16.7046  &  8.5706\\
 & 2000 &   & 0.4900   &14.6124 &3.6531 & 34.3767  & 19.5655 \\
 &  500    &   & 0.4750  &20.4632  & 5.1158  &7.9367  & 5.1953 \\
 & 1000  & $50\%$   & 0.2800    &18.7144  &4.6786 &16.7158 &  13.1580 \\
 & 2000  &  & 0.1850 & 23.2432 &5.8108  &34.4379  & 30.2443\\
\hline
\end{tabular}%
\end{center}
\end{table}

\clearpage
\begin{table}[htb]
 \begin{center}
\caption{\label{tab:ex3.3}  Simulation results of Example 3, scenario (iii). }
\vspace{0.1in}
\resizebox*{!}{\dimexpr\textheight-2\baselineskip\relax}
{
\begin{tabular}{ cccccccc }
\hline
 Method &$p$ &  $R^2$& CP & AFP & AFN & AMS & Time3 (s)\\
 \hline
 FR &500 &  &  0.9650  & 0.5300   & 0.0350   &  5.4950 & 0.6673\\
    &1000& $90\%$ &  0.9000  &0.7400   & 0.1000  &   5.6400    & 1.3855\\
    &2000&  &  0.8750  &0.8700   & 0.1250  &   5.7450  &    2.8385 \\
    &500 &  & 0.1100  & 0.9500   & 0.8950   &    5.0550  & 0.5832 \\
    &1000& $70\%$ & 0.1150  & 1.3450   & 0.8850   &    5.4500   & 1.3032 \\
    &2000&  &  0.0600  &  1.6650   & 0.9600 &    5.7050   &     2.5906\\
    &500 & & 0.0150 & 1.3250  &  2.2550   &   4.0700 & 0.4522\\
    &1000&  $50\%$ &  0.0150   &  1.5000   & 2.8300    &     3.6700  &  0.7955 \\
    &2000&  & 0.0050  &  1.3750    &3.2850    &    3.0900    &1.2730\\
\\
GFR($J=2$) &500 &  & 0.7550  & 2.0550  &  0.2450   &       6.8100 &  0.3863 \\
    &1000& $90\%$ &  0.6450  & 2.0550  &  0.3550   &     6.7000 & 0.7515\\
    &2000&  &  0.6050  & 2.1850    &0.3950 &         6.7900  &  1.5292 \\
    &500 &  & 0.2000  & 1.8900  &  0.8000 &    6.0900  &  0.3454\\
    &1000& $70\%$ &  0.1550 & 1.9800  &  0.8500    &      6.1300&      0.7010 \\
    &2000&  &  0.1150 &2.0700   & 0.8900   &     6.1800  &   1.4215 \\
    &500 &  & 0.0750   &  1.7300   & 1.4400   &  5.2900 & 0.2807 \\
    &1000& $50\%$ &  0.0400  & 1.6800   & 1.6700    &      5.0100 &   0.5192\\
    &2000&  &0.0200  & 1.5950   & 2.0050   &        4.5900 &     0.9061 \\
\\
GFR($J=4$)&500 &  & 0.2850  & 4.1350   & 0.7150    &      8.4200 & 0.2630 \\
    &1000& $90\%$ &  0.2200  & 4.1200  &  0.7800   &        8.3400&  0.5168 \\
    &2000&  &   0.2050  & 4.1550   & 0.7950  &     8.3600&1.0404 \\
    &500 &  &  0.1700   &3.9050   & 0.9250 &   7.9800  &  0.2480 \\
    &1000& $70\%$ &  0.1300  & 3.8550   & 1.0550    &  7.8000& 0.4766 \\
    &2000&  &  0.1000 & 3.8050   & 1.1450    &    7.6600 &     0.9367 \\
    &500 &  &  0.0900  &  3.3000  &  1.4600   &      6.8400 & 0.1922 \\
    &1000& $50\%$ & 0.0600  &   3.2950   & 1.6350   &   6.6600 &       0.3689 \\
    &2000&  & 0.0300  & 3.2500  &  1.7500   &    6.5000&        0.7137 \\
\\
SIS &500 &  &0  & 25.4200   & 1.4200 &  29.0000 &0.0213 \\
    &1000& $90\%$ &  0   & 25.5600   & 1.5600& 29.0000 & 0.0765 \\
    &2000&  & 0    &25.6300   & 1.6300&   29.0000 & 0.2909 \\
    &500 &&  0     &  25.5400   & 1.5400&29.0000&0.0227\\
    &1000&  $70\%$  & 0   &25.6750    &1.6750 &29.0000  & 0.0811 \\
    &2000&  & 0    & 25.7700   & 1.7700& 29.0000 &  0.3040 \\
    &500 &  & 0    & 25.7200  &  1.7200& 29.0000  &  0.0227 \\
    &1000& $50\%$ &  0  & 25.8750   & 1.8750 & 29.0000 &0.0769 \\
    &2000&  &0    &25.9950    &1.9950&    29.0000 &0.3074 \\
\\
ISIS &500 &  &0   & 112.0000  &  1.0000 & 116.0000  &0.0549 \\
    &1000& $90\%$ &  0   & 112.0150    &1.0150 & 116.0000  & 0.1387 \\
    &2000&  & 0   &  112.0150    &1.0150 &116.0000 & 0.4183 \\
    &500 &&  0      & 112.0750   & 1.0750 &  116.0000 &0.0554\\
    &1000&  $70\%$  & 0    & 112.1900   & 1.1900 &116.0000  & 0.1313 \\
    &2000&  & 0    &  112.3100    &1.3100  &116.0000 &  0.4002 \\
    &500 &  & 0     & 112.2000    &1.2000& 116.0000  &  0.0515 \\
    &1000& $50\%$ &  0  & 112.4300    &1.4300& 116.0000 &0.1294 \\
    &2000&  &0      & 112.5650    &1.5650 & 116.0000&0.4236 \\

\hline
\end{tabular}%
}

\end{center}
\end{table}



\clearpage
\begin{table}[htb]
 \begin{center}
 \caption{\label{tab:271gene}  Indices of selected genes for the breast cancer data.}
\vspace{0.1in}
\begin{tabular}{ cc }
\hline
  Method &   Genes  \\
 \hline
 FR  &                       272, 167, 5342     \\
  GFR($J=2$)       &  272, 166 \\
  GFR($J=4$)  &                   272, 166, 275, 267, 24032, 11913, 11870, 17439   \\
\hline
\end{tabular}%
\end{center}
\end{table}

\begin{table}[htb]
 \begin{center}
 \caption{\label{tab:271pmse}  Average PMSEs of different methods when applied to the breast cancer data.}
\vspace{0.1in}
\begin{tabular}{ cccc c}

\hline
SIS	&ISIS&	FR	&GFR($J=2$)&	GFR($J=4$)\\
\hline
0.4804&0.5260&0.4807&0.5081&0.4365\\
\hline
\end{tabular}
\end{center}
\end{table}

\clearpage
\begin{table}[htb]
 \begin{center}
 \caption{\label{tab:ratgene}  Selected probes for the rats eye data.}
\vspace{0.1in}
\begin{tabular}{ cc }
\hline
  Method &   Genes  \\
 \hline
 FR  &                       1383110\_at, 1389584\_at, 1392692\_at \\
  GFR($J=2$)       &  1383110\_at, 1389584\_at, 1392692\_at, 1378099\_at \\
  GFR($J=4$)  &                   1383110\_at, 1389584\_at, 1383673\_at, 1386683\_at   \\
\hline
\end{tabular}%
\end{center}
\end{table}

%

\begin{table}[htb]
 \begin{center}
 \caption{\label{tab:ratpmse}  Average PMSEs of different methods when applied to the rats eye data.}
\vspace{0.1in}
\begin{tabular}{ cccc c}
\hline
SIS	&ISIS&	FR	&GFR($J=2$)&	GFR($J=4$)\\
\hline
0.6291&	0.8502&	0.6948	&0.6592&	0.6026\\
\hline
\end{tabular}
\end{center}
\end{table}

\end{document}